# The "Academic Trace" of the Performance Matrix: A Mathematical Synthesis of the h-Index and the Integrated Impact Indicator (I3)

Fred Y. Ye [a*] and Loet Leydesdorff [b]

[a] School of Information Management, Nanjing University, Nanjing 210093, CHINA
*yye@nju.edu.cn

[b] Amsterdam School of Communication Research (ASCoR), University of Amsterdam, Kloveniersburgwal 48, 1012 CX Amsterdam, The Netherlands
loet@leydesdorff.net

**Abstract**

The h-index provides us with nine natural classes which can be written as a matrix of three vectors. The three vectors are: **X**=($X_1$, $X_2$, $X_3$) indicate publication distribution in the h-core, the h-tail, and the uncited ones, respectively; **Y**=($Y_1$, $Y_2$, $Y_3$) denote the citation distribution of the h-core, the h-tail and the so-called "excess" citations (above the h-threshold), respectively; and **Z**=($Z_1$, $Z_2$, $Z_3$)= ($Y_1$-$X_1$, $Y_2$-$X_2$, $Y_3$-$X_3$). The matrix **V**=$(X,Y,Z)^T$ constructs a measure of academic performance, in which the nine numbers can all be provided with meanings in different dimensions. The "academic trace" tr(**V**) of this matrix follows naturally, and contributes a unique indicator for total academic achievements by summarizing and weighting the accumulation of publications and citations. This measure can also be used to combine the advantages of the h-index and the Integrated Impact Indicator (I3) into a single number with a meaningful interpretation of the values. We illustrate the use of tr(**V**) for the cases of two journal sets, two universities, and ourselves as two individual authors.

**Keywords**: performance matrix; academic trace; publications; citations; I3; h-index; h-core; h-tail

## 1. Introduction

Since Garfield (1955) introduced the Science Citation Index (SCI) and suggested the Impact Factor (IF) as an important indicator, citation analysis has increasingly become a scientific field of studies (Garfield, 1979). Developed by scientometricians, data around SCI and IF have been the subject of many studies. However, the skewness of citation and publication distributions delegitimizes the use of averages (Seglen, 1992, 1997; cf. Rousseau & Leydesdorff, 2011).

As a non-parametric alternative, Bornmann & Mutz (2011) suggested to turn to the six percentile rank classes in use by the National Science Foundation (NSF) in the *Science & Engineering Indicators* (NSB, 2012): top-1%, top-5%, top-10%, etc., of highly-cited papers. Leydesdorff *et al.* (2011) developed the Integrated Impact Indicator (I3) that is based on normalization in terms of percentile ranks of the distribution. More recently, the top-10% of publications in terms of citations has increasingly been used as an “excellence indicator” in university rankings (e.g., Bornmann et al., 2012; Waltman *et al.*, 2012).

Rousseau (2012) studied the relation between I3—that is, ranking based on the integration of weighted percentile rank classes—with the h-index (Hirsch, 2005)—that is, ranking based on a core-tail concept. Can the core, the tail, and the uncited papers be considered as three relevant classes for I3? He concluded that although “the h-index can be written in such a way that it formally looks like an I3 score, it is *not* an I3 score. The reason is that the scores $x_k$ and the classes may not depend on the set A.” In other words: the number of documents in the h-core (and h-tail, respectively) would determine the weight of the class, whereas these two numbers are independent from the number of documents in the sample using percentile ranks across a distribution.

Furthermore, the h-index uses the document set under study as its own reference set, whereas I3 ranks the document set as a sample against a (larger) reference set. For example, one can rank the I3-value of a journal among other (similar) journals (Leydesdorff & Bornmann, 2011) or one university among other ones (Bornmann et al., 2013; Leydesdorff & Shin, 2011; Prathap & Leydesdorff, 2012; Waltman *et al.*, 2012) using the superset of similar samples for the reference.

The h-index is based on publications and citations, and was introduced in 2005 (Hirsch, 2005). Its simplicity has made it attractive for use in academic performance measurement (Alonso *et al.*, 2009; Egghe, 2010) and has led to a meaningful unification of publications and citations (Ye, 2011). Yet, it could be shown that the h-index is logically flawed in the sense that (in a static time window) it is not independent (Marchant, 2009) and not consistent (Waltman & van Eck, 2012). However, this is not an issue in dynamic cases (Ye, 2012).

In the meantime, the h-index and h-type indicators were further developed and improved. Kuan *et al.* (2011a), for example, proposed the c-descriptor and t-descriptor for analyzing patent performance of assignees. Kuan et al. (2011b) used the h-core and h-tail centroids, which are located at the geometric centers of the h-core and h-tail areas (Chen *et al.*, 2013). Zhang (2013a & b) developed a novel triangle mapping technique and introduced the h' index where citations in h-tail are considered as negative contribution, while Thor & Bornmann (2011) introduced indices h-upper, h-lower, h-center with a web application based on Google Scholar.

In this study, we propose to combine the ideas of I3 with the h-index. This will lead us to a new academic vector metrics, characterized by a set of three vectors (**X**, **Y**, **Z**) that can be combined into the performance matrix **V**, and then summarized as the trace of this matrix: tr(**V**). The various elements of both this matrix and the trace will be provided with detailed interpretations for the case of academic publications and citations, but the reasoning is more abstract and can further be elaborated for other applications (e.g., patents or, more generally, any skewed distribution).

## 2. Methodology

On the basis of the definition (Leydesdorff & Bornmann, 2011; Leydesdorff *et al.*, 2011; Rousseau & Ye, 2012a), one can formalize I3 as follows:

$$I3 = \sum_{i=1}^{C} f(X_i) \cdot X_i \tag{1}$$

where $X_i$ indicates the percentile ranks and $f(X_i)$ denotes the frequencies of the ranks with $i$=[1,C] as the percentile rank classes. In other words, the measures $X_i$ are divided into C classes each with a scoring function $f(X_i)$ or weight ($w_i$) so that one can aggregate as follows:

$$I3 = \sum_{i} w_i X_i \tag{2}$$

More generally, when one ranks publications according to their citations from high to low, one obtains a C-P rank distribution—the citation curve—as shown in Figure 1. We added the three sections that are relevant for the h-index: the h-core, h-tail, and the uncited (zero citations) publications ($P_z$) respectively (Ye & Rousseau, 2010; Chen *et al.*, 2013; Liu *et al.*, 2013). Furthermore, Zhang (2009) proposed to call the area above the h-core a representation of "excess citations," that is, citations which are gathered, but do not further contribute to the h-value.

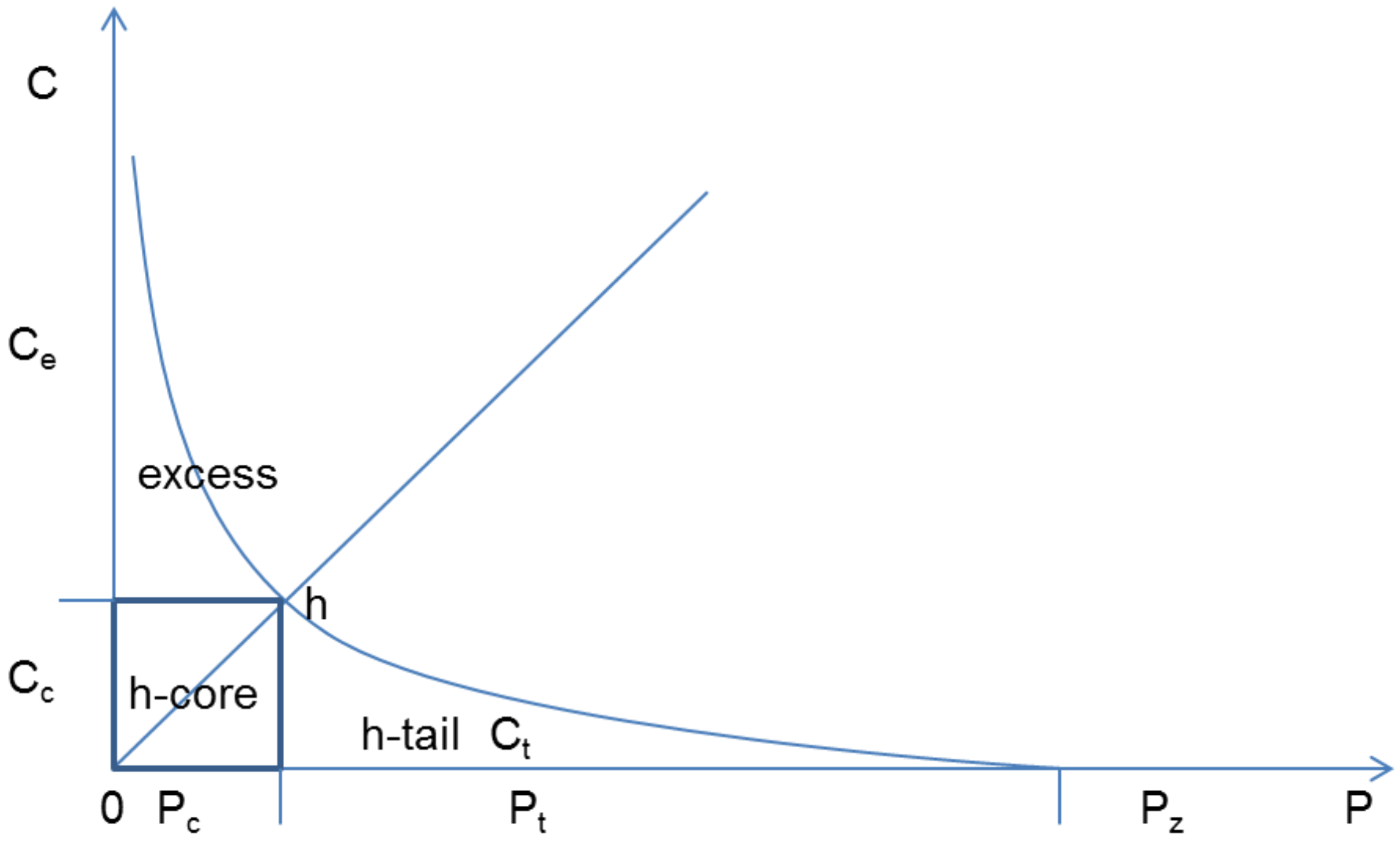

**Fig. 1**: The rank distribution of citations versus publications. The different domains in relation to the *h*-index are indicated.

## *2.1 The Performance Matrix*

Similarly to I3 (in Eq. 2), one can define a weighted I3-like measure corresponding to publications and citations in the core-tail framework of Figure 1, and formulate I3-like indicators for both publications (I3X) and citations (I3Y) based on the three respective classes, as follows:

$$I3X = x_c P_c + x_t P_t + x_z P_z = \frac{P_c}{P_c + P_t + P_z} \cdot P_c + \frac{P_t}{P_c + P_t + P_z} \cdot P_t + \frac{P_z}{P_c + P_t + P_z} \cdot P_z \quad (3)$$

$$I3Y = y_c C_c + y_t C_t + y_e C_e = \frac{C_c}{C_c + C_t + C_e} \cdot C_c + \frac{C_t}{C_c + C_t + C_e} \cdot C_t + \frac{C_e}{C_c + C_t + C_e} \cdot C_e \quad (4)$$

In Eq. (3), $P_c$ = h denotes the number of publications in the *h*-core, $P_t$ the number of publications in the *h*-tail, $P_z$ the number of uncited (zero citation) publications. In Eq. (4), $C_c = h^2$, $C_t$ the number of citations in the *h*-tail, $C_e$ the number of citations in the excess area (Zhang, 2009), $C_h = C_c + C_e$ indicates the total number of citations in the h-core. $P= P_c+P_t+ P_z$ is the total number of publications and $C = C_c+ C_t + C_e$ is the total number of citations, with the following relation between them:

$$C_c = P_c^2 = C_h - C_e = h^2 \quad (5)$$

Our scheme of weighting scores (following Eqs. 3 and 4) can be considered as follows: $x_c=P_c/(P_c+P_t+P_z)$, $x_t=P_t/(P_c+P_t+P_z)$, $x_z=P_z/(P_c+P_t+P_z)$, $y_c=C_c/(C_c+C_t+C_e)$, $y_t=C_t/(C_c+C_t+C_e)$ and $y_e=C_e/(C_c+C_t+C_e)$ given that $P_c$, $P_t$, $P_z$, $C_c$, $C_t$ and $C_e$ measure each three classes. Analogous to the scoring function in I3, we apply weights in the case of I3X and I3Y as follows: $x_c + x_t + x_z =1$ and $y_c + y_t + y_e =1$. Since $I3X=x_cP_c+x_tP_t+x_zP_z$ and $I3X=y_cC_c+y_tC_t+y_eC_e$, it follows that $x_c + x_t + x_z =1$ and $y_c + y_t + y_e =1$.

However, one can make the classes relative to—that is, equivalent to a percentage of—the size of the sets under study (cf. Rousseau, 2012) by normalizing to fractions (percentages), as follows:

$$X_1 = \frac{P_c}{P_c + P_t + P_z} \cdot P_c = \frac{P_c^2}{P} \tag{6}$$

$$X_2 = \frac{P_t}{P_c + P_t + P_z} \cdot P_t = \frac{P_t^2}{P} \tag{7}$$

$$X_3 = \frac{P_z}{P_c + P_t + P_z} \cdot P_z = \frac{P_z^2}{P} \tag{8}$$

$$Y_1 = \frac{C_c}{C_c + C_t + C_e} \cdot C_c = \frac{C_c^2}{C} \tag{9}$$

$$Y_2 = \frac{C_t}{C_c + C_t + C_e} \cdot C_t = \frac{C_t^2}{C} \tag{10}$$

$$Y_3 = \frac{C_e}{C_c + C_t + C_e} \cdot C_e = \frac{C_e^2}{C} \tag{11}$$

These six numbers $X_1$, $X_2$, $X_3$,$Y_1$, $Y_2$ and $Y_3$ can be re-organized and defined as two independent vectors as follows:

$$X = (X_1, X_2, X_3) = (P_c^2 / P, P_t^2 / P, P_z^2 / P) \tag{12}$$

$$Y = (Y_1, Y_2, Y_3) = (C_c^2 / C, C_t^2 / C, C_e^2 / C) \tag{13}$$

The vector measures **X**=($X_1$, $X_2$, $X_3$) and **Y**=($Y_1$, $Y_2$, $Y_3$) indicate the distributions of publications and citations in the h-core, h-tail and the uncited areas, respectively.

Since C–P is the difference between the total number of citations and the total number of publications and it no longer indicates the rank distribution as a simple consistent measure (Rousseau & Ye, 2011), one is allowed to derive the additional vector **Z**, as follows:

$$Z = (Z_1, Z_2, Z_3) = (Y_1 - X_1, Y_2 - X_2, Y_3 - X_3) \quad (14)$$

It follows logically that **Z** is also a consistent measure. However, the vector **Z** can be provided with an interpretation beyond its generation as an arithmetic subtraction of the number of publications from the number of citations. The terms of **Z** can be appreciated as the fraction of citations (with the dimensionality of citation [$C_i^2/C$]) minus the fractions of publications (with the dimensionality of publication [$P_i^2/P$]), so that Z is a set of meaningful indicators, where $C_i$ covers $C_c$, $C_t$ and $C_e$, and $P_i$ includes $P_c$, $P_t$ and $P_z$. $Z_3$ (= $C_e^2/C - P_z^2/P$) is a complex indicator because one considers the excess citations as a possible compensation for the uncited publications. The fraction of uncited publications contributes negatively to $Z_3$, but this can be compensated by the fraction excess citations in a set.

Let us call **X**, **Y** and **Z** academic vectors, consisting of {$X_i$, $Y_i$, $Z_i$} (i=1, 2, 3). On this basis, one can construct a unique matrix **V** for measuring the total distribution of academic achievements as follows:

$$V = \begin{pmatrix} X_1 & X_2 & X_3 \\ Y_1 & Y_2 & Y_3 \\ Z_1 & Z_2 & Z_3 \end{pmatrix} = \begin{pmatrix} X \\ Y \\ Z \end{pmatrix} = \begin{pmatrix} X & Y & Z \end{pmatrix}^T \quad (15)$$

The matrix **V** contains nine numbers ($X_1$, $X_2$, $X_3$; $Y_1$, $Y_2$, $Y_3$; $Z_1$, $Z_2$, $Z_3$), but only the two vectors (**X, Y**) are independent. Therefore, the matrix **V** provides us with a two-dimensional measure of academic achievements,[1] with the following specific meanings:

- $X_1$, $X_2$ and $X_3$ indicate the publication distributions in the h-core, h-tail, and the uncited areas, respectively;
- $Y_1$, $Y_2$ and $Y_3$ denote the citation distribution of the h-core, h-tail, and the excess areas, respectively;
- Column vector ($X_1$, $Y_1$, $Z_1$) indicates the distribution of publications and citations in the h-core;
- Column vector ($X_2$, $Y_2$, $Z_2$) does so in the *h*-tail, whereas row vector ($X_1$, $X_2$, $X_3$) denotes the publication view and row vector ($Y_1$, $Y_2$, $Y_3$) the citation view;
- Column vector ($X_3$, $Y_3$, $Z_3$) reflects the uncited publications, the excess citations, and the difference between these two fractions; row vector ($Z_1$, $Z_2$, $Z_3$) provides citations minus publications, where $X_3$ marks the fraction of uncited publications and $Y_3$ the fraction of excess citations, and $Z_3$ indicates their corresponding differences.

[1] There are *maximally* two independent dimensions because there can also be a relation between the number of publications and citations.

Furthermore, the number pair ($X_1$, $Y_1$)=($h^2/P$, $h^2/C$) reflects achievements in the h-core and ($X_2$, $Y_2$) in the h-tail. Larger values of $X_i$ and $Y_i$ (i=1, 2, 3) indicate academic achievements. However, $X_3$ can be excluded as an indicator of achievement when the number of uncited publications is not considered as adding to the rank. (Alternatively, one can set $X_3$ to zero in an evaluation.) Except for $X_3$, the larger values of {$X_i$, $Y_i$, $Z_i$} are, the higher always the level of academic achievement indicated.

Among the above nine numbers, it may look as if there are six independent ones ($X_1$, $X_2$, $X_3$; $Y_1$, $Y_2$, $Y_3$). However, there are the following linear relations among them: $P_t$ =P- $P_c$ - $P_z$, $P_c$ = h, $C_c$ = $h^2$, $C_t$ = C-$C_h$,and $C_e$ = $C_h$-$C_c$. Consequently one needs only five really independent numbers (P, $P_z$, h, C, $C_h$) in the data collection for the calculation of the performance matrix

*2.2 The Academic Trace*

The trace of matrix **V**—we propose to call it "the academic trace" of the "performance matrix"—provides us with a single and unique measure that summarizes academic achievements as follows:

$$T = tr(V) = X_1 + Y_2 + Z_3 \qquad (16)$$

The trace T of the matrix **V** is a mathematical result that follows naturally from the above core-tail framework of the h-index when combined with the idea of relative frequencies used for I3. When we classify $X_i$, $Y_i$, and $Z_i$ in the core-tail plane, we obtain matrix **V**, which provides us with the trace T of this matrix. For each academic source (in a specific database and time span), its **V** and T are determinate and unique. The trace of **V** can be specified as $T=X_1+Y_2+Z_3=h^2/P+C_t^2/C+(C_e^2/C-P_z^2/P)$, and thus covers the distributions of the h-core, the h-tail, and the uncited areas. This summarizes the representative information distributed over the e-area, the h-area, the t-area, and the uncited area. In other words, T = tr (**V**) provides a scalar number that can be used as an indicator that summarizes **V** as a matrix consisting of three vectors **X, Y** and **Z**.

The three components of T are themselves meaningful: $X_1$= $P_c^2$ /P=$h^2$/P indicates publications in the h-core and is determined by h when P is a constant. In other words, $X_1$ is a normalized publication score for the h-core. $Y_2$= $C_t^2$ /C, provides a normalized citation measure for the h-tail, which represents citations in the h-tail. In our opinion, citations in the h-tail should not be considered as meaningless, but can be added to the achievements measured by $X_1$. When a new paper is first published and begins to earn citations, for example, these citations always fall initially in the h-tail. But when the citations to a publication accumulate, these same publications and citations may be included in the h-core in a later stage.

Finally, $Z_3 = C_e^2/C - P_z^2/P$, measures the difference between the fraction of excess citations and the fraction of uncited publications in terms of the respective areas in Figure 1, so that the additional impact of excess citation in the set can compensate for uncited publication. Note that this value can also be negative when the number of uncited publications is larger than the sum of the excess citations.

$X_1$, $Y_2$ and $Z_3$ thus construct a synthetic measure of the h-core, the h-tail, the excess and uncited areas, so that $T = X_1+Y_2+Z_3$ reflects the publication-citation distribution in the core-tail plane of Figure 1. However, the trace is not just a number: the trace can be provided with a meaningful interpretation in terms of the various areas in the citation distribution of Figure 1. Thus, T can be considered as a synthetic indicator for measuring total academic achievements, which provides a total number of academic historical records. The larger the value of T, the more academic accumulation is measured.

While the value of the h-index marks only a single cut-off level in the core-tail plane, the value of T includes summary information across the h-core, h-tail, and the uncited areas. For this reason, T provides an improved h-index (which can be extended to similar measures such as the g-index, etc.). Compared with the h-index, the academic trace T seems more complex, but is nevertheless also an indicator that is simple in the computation.

Note that T is sensitive to increases or decreases in the performance, while the h-index can only increase. In our opinion, a newly added, but yet uncited publication can first meaningfully decrease T. All cited publications in the h-tail or excess citations to the h-core, however, lead to increases of T. As with the h-index, the analyst may wish to limit the time window for both publications and their citations when comparing sets for the evaluation.

The academic vectors, matrices, and traces can be applied to information sources at different levels, including scholars, research groups, journals, institutions, universities, countries, and even topics. However, one should always compare "like with like" (Martin & Irvine, 1983), in the same field. (One needs a scaling normalization when comparing among different fields). Since our reasoning is abstract, *both papers and patents can be used as data sources.* Of course, all values of academic vectors, matrices, and traces remain contingent upon the databases and time windows used for the data collection.

## 3. Data

### *3.1 Journals*

In order to compare our results with Leydesdorff & Bornmann (2011)'s initial study about I3, we collected the following journal data as samples: (1) journals in information science and library science recorded in the JCR for SSCI, and (2) *Nature, Science* and *PNAS* as leading interdisciplinary journals recorded in the JCR of SCI. The data is provided in an Appendix. All data were downloaded from Web of Science (WoS, updated on March 22, 2013), with a two-year time window (2009-2010) and a five-year time window (2008-2012) respectively, in order to make comparisons with IF and IF5 also possible. On the basis of this data, all indicators can be computed.

### *3.2. Other units of analysis*

In order to show the general applicability of this trace-measure, we also provide examples at different levels: for academics (ourselves, using 10 years of data from 2003-2012 in WoS) and for the comparison between two German universities, using 2012 data in WoS. (Table A2 in the Appendix shows this data.) We chose ourselves in order to avoid privacy issues and these two universities because their names are not ambiguous. The use of the indicator, of course, can be scaled up.

## 4. Results

### *4.1 Journals*

On the basis of Eqs. (6) to (16), and using the data in the Appendix, we calculated results as shown in Table 1.

**Table 1:** Selected journals results; the top-3 among the multidisciplinary journals and the top-20 ranked highest in LIS, according to the trace value T.

| Journal | $X_1$ | $X_2$ | $X_3$ | $Y_1$ | $Y_2$ | $Y_3$ | $Z_1$ | $Z_2$ | $Z_3$ | T |
|---|---|---|---|---|---|---|---|---|---|---|
| **Multidisiciplinary journals** | | | | | | | | | | |
| 1. *PNAS* | 1.57 | 7492 | 16.4 | 822.5 | 176584 | 149.9 | 820.9 | 169092 | 133.5 | 176719 |
| 2. *Nature* | 7.2 | 1871 | 657 | 7440 | 74359 | 4683 | 7433 | 72488 | 4026 | 78392 |
| 3. *Science* | 5.9 | 2274 | 411 | 5356 | 72483 | 3266 | 5350 | 70208 | 2855 | 75344 |
| **Library and Information Sciences** | | | | | | | | | | |
| 1. *Scientometrics* | 0.68 | 310 | 4.76 | 37.17 | 1461.2 | 9.486 | 36.49 | 1151.2 | 4.722 | 1466.6 |
| 2. *J Am Soc Inf Sci Tec* | 0.82 | 222.3 | 39.1 | 66.56 | 1190.9 | 40.49 | 65.73 | 968.61 | 1.388 | 1193.1 |
| 3. *J Am Med Inform Assn* | 2.33 | 157.1 | 2.74 | 147.9 | 915.51 | 24.41 | 145.6 | 758.39 | 21.68 | 939.52 |
| 4. *J Health Commun* | 0.75 | 106 | 7.48 | 26.72 | 399.81 | 7.249 | 25.98 | 293.85 | -0.23 | 400.32 |
| 5. *Int J Geogr Inf Sci* | 0.97 | 96.57 | 5.85 | 35.93 | 386.06 | 6.521 | 34.96 | 289.49 | 0.669 | 387.69 |
| 6. *J Informetr* | 3.09 | 64.04 | 0.24 | 92.73 | 275.06 | 55.21 | 89.65 | 211.02 | 54.97 | 333.12 |
| *7. MIS Quart* | 4.55 | 43.68 | 0.41 | 141.2 | 274.81 | 27.03 | 136.7 | 231.13 | 26.62 | 305.98 |
| *8. J Knowl Manag* | 0.76 | 88.36 | 1.48 | 18.25 | 257.99 | 9.46 | 17.49 | 169.62 | 7.975 | 266.72 |
| *9. Inform Manage-Amster* | 1.71 | 61.45 | 0.65 | 43.14 | 255.17 | 10.16 | 41.44 | 193.71 | 9.511 | 266.39 |
| *10. Int J Inform Manage* | 0.76 | 44.38 | 25.8 | 25.73 | 257.8 | 7.425 | 24.97 | 213.42 | -18.3 | 240.23 |
| *11. Telecommun Policy* | 0.76 | 64.61 | 6.42 | 21.05 | 226.49 | 4.651 | 20.29 | 161.88 | -1.77 | 225.49 |
| *12. Gov Inform Q* | 1.05 | 51.43 | 20.2 | 44.49 | 216.71 | 15.58 | 43.44 | 165.28 | -4.6 | 213.16 |
| *13. Inform Process Manag* | 0.67 | 59.71 | 6.02 | 15.19 | 184.08 | 11.02 | 14.52 | 124.37 | 4.996 | 189.75 |
| *14. J Comput-Mediat Comm* | 1.55 | 54.2 | 1.08 | 35.57 | 153.35 | 33.62 | 34.02 | 99.142 | 32.54 | 187.44 |
| *15. J Inf Sci* | 0.84 | 53.44 | 2.64 | 16.78 | 182.32 | 4.729 | 15.94 | 128.88 | 2.09 | 185.25 |
| *16. Inform Syst Res* | 1.94 | 47.08 | 1.15 | 49.93 | 157.34 | 18.55 | 47.99 | 110.26 | 17.4 | 176.68 |
| *17. Eur J Inform Syst* | 0.82 | 64.65 | 1.01 | 17.98 | 156.5 | 5.548 | 17.16 | 91.849 | 4.538 | 161.85 |
| *18. J Manage Inform Syst* | 1.12 | 37.8 | 4.96 | 27.17 | 108.7 | 12.57 | 26.05 | 70.898 | 7.61 | 117.43 |
| *19. J Med Libr Assoc* | 0.38 | 33.98 | 43.5 | 14.27 | 155.13 | 0.502 | 13.9 | 121.15 | -43 | 112.5 |
| *20. J Doc* | 0.37 | 30.8 | 28.9 | 8.795 | 129.47 | 4.747 | 8.426 | 98.668 | -24.2 | 105.68 |

These results show that the trade-off is multi-dimensional and different choices therefore are possible in the comparisons. For example, the *Journal of Informetrics* (JoI) has a value for $X_1$ (that is, the normalized indicator of core-h publications) higher than the *Journal of the American Society for Information Science and Technology* (JASIST), but JASIST has a much larger value than JoI in the h-tail, and eventually in the total accumulation as reflected by the trace. The academic matrices and traces for these two journals are:

$$V_{JOI} = \begin{pmatrix} 3.09 & 64.04 & 0.24 \\ 92.73 & 275.06 & 55.21 \\ 89.65 & 211.02 & 54.97 \end{pmatrix} \quad (17)$$

$$tr(V_{JOI}) = 3.09 + 275.06 + 54.97 = 333.12 \quad (18)$$

$$V_{JASIST} = \begin{pmatrix} 0.82 & 222.3 & 39.1 \\ 66.56 & 1190.9 & 40.49 \\ 65.73 & 968.61 & 1.388 \end{pmatrix} \quad (19)$$

$$tr(V_{JASIST}) = 0.82 + 1190.9 + 1.388 = 1193.1 \quad (20)$$

Using the academic trace, *JASIST* is thus indicated as higher in terms of total performance than JoI (as well as *MIS Quarterly*). However, the trace (Eq. 20) shows that JASIST earns 99.8% (= 1190.9/1193.1) of its performance credit by citations in the h-tail ($Y_2$). These results completely accord with those for I3 of Leydesdorff and Bornmann (2011). The results of the comparison among the multidisciplinary journals (*PNAS* with higher value of *T* than *Nature*) are also similar to the ones for I3 reported by these authors.

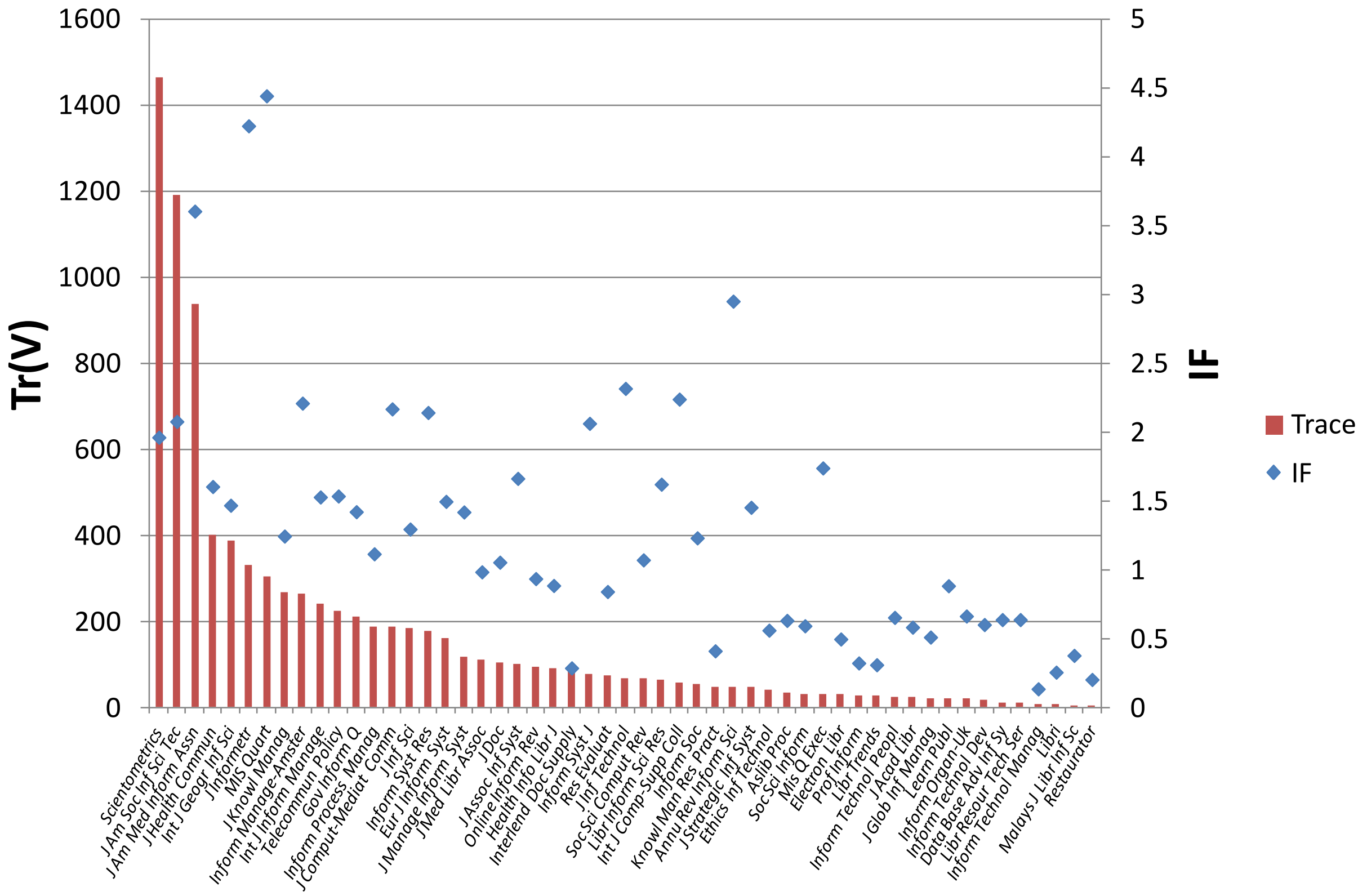

**Fig. 2**: Journal rank by academic traces (2 years) compared with two-year impact factors (IF).

Figure 2 shows the relations between the trace-values and two-year impact factors for the journals in the WoS Subject Category "Library and Information Sciences" (insofar as T > 0). This figure is not essentially different using five years instead of two. Table 2 provides the Pearson and rank-order correlations.

**Table 2**: Correlations between values of traces and impact factors for both two- and five-year time windows (IF and IF5 from JCR 2011).

| Correlations | | Spearman (2-tailed) | | | |
|---|---|---|---|---|---|
| | | T | T5 | IF | IF5 |
| Pearson (2-tailed) | T | | 1.000** | .804** | .812** |
| | T5 | .994** | | .802** | .812** |
| | IF | .234* | .600** | | .665** |
| | IF5 | .506** | .527** | .568** | |

* $p < .05$ ; ** $p < .01$

In our opinion, the strong reduction of the complexity of the citation curve into a single number using the h-index or IF have been unfortunate choices. In the case of IF, an average is taken over a very skewed distribution. The h-index is non-parametrical, but the size of the samples influences the attribution into a classificatory scheme: larger and older sets tend to have larger h-values for no other reason than the accumulative effect of having grown older and larger.

An additional normalization is therefore proposed by us in the case of the vectors **X, Y**, and **Z**, analogously to using percentiles for I3. The choice for the number of percentile rank classes (such as six or hundred) was hitherto conventional (Bornmann & Mutz, 2011; Leydesdorff et al., 2011; NSB, 2012). Our new measure offers a reasoned reduction of the complexity from first nine possible to three independent classes that can be aggregated as a mathematically defined trace.

## *4.2 Other examples*

The trace can be measured for any download from WoS. For example, one can rank institutional units such as universities. Table 3 provides a comparison between the German universities of Heidelberg and Hamburg and Table 4 extends the analysis to individual authors using our own track records as a (harmless) example.

**Table 3**: Two German universities compared using WoS data 2012.

| | $X_1$ | $X_2$ | $X_3$ | $Y_1$ | $Y_2$ | $Y_3$ | $Z_1$ | $Z_2$ | $Z_3$ | T |
|---|---|---|---|---|---|---|---|---|---|---|
| Univ Heidelberg | 0.0935 | 506.26 | 2103 | 37.257 | 3418 | 59.009 | 37.163 | 2911.8 | -2044 | 1374.03 |
| Univ Hamburg | 0.1852 | 232.39 | 811 | 40.917 | 1184.1 | 244.25 | 40.732 | 951.71 | -566.5 | 617.836 |

Note that in the setting of a large university $Z_3$ will often be negative because uncited publications ($X_3$) can be expected to prevail and excess citations ($Y_3$) may be scarce. Although the University of Hamburg is lower on both these partial indicators, the University of Heidelberg is on the aggregate more than twice as large in terms of the value of the trace, mainly because of the high value of $Y_2$ that indicates citations to the publications in the tail.

As a second extension, we use our own records because the issue of measuring authors at the individual level involves sometimes a privacy issue.

**Table 4**: Academic matrices and traces for the two authors during the period 2003-2012.

| Author | $X_1$ | $X_2$ | $X_3$ | $Y_1$ | $Y_2$ | $Y_3$ | $Z_1$ | $Z_2$ | $Z_3$ | T |
|---|---|---|---|---|---|---|---|---|---|---|
| Leydesdorff L | 5.1702 | 58.73 | 3.75 | 243.45 | 332.53 | 166.01 | 238.28 | 273.8 | 162.26 | 499.956 |
| Ye FY | 1 | 4.84 | 3.24 | 8.6806 | 6.125 | 9.3889 | 7.6806 | 1.285 | 6.1489 | 13.2739 |

$$V_{Leydesdorf} = \begin{pmatrix} 5.17 & 58.73 & 3.75 \\ 243.45 & 332.53 & 166.01 \\ 238.28 & 273.8 & 162.26 \end{pmatrix} \quad (21)$$

$$tr(V_{Leydesdorf}) = 5.17 + 332.53 + 162.26 = 499.96 \quad (22)$$

$$V_{Ye} = \begin{pmatrix} 1 & 4.84 & 3.24 \\ 8.68 & 6.13 & 9.39 \\ 7.68 & 1.29 & 6.15 \end{pmatrix} \quad (23)$$

$$tr(V_{Ye}) = 1 + 6.13 + 6.15 = 13.27 \quad (24)$$

Using individual authors, the samples may be too small for other statistics. By using the trace, however, all absolute numbers (see Appendix) are normalized to relative ones by using Eqs. (6-11).

## 5. Analysis

Suppose that the C-P rank distribution were a continuous function C(x), where x denotes the publications ranked by citations. We can then generalize $C_e$, $C_c$ and $C_t$ as follows ($P_c$=h)

$$C_h = C_e + C_c = \int_1^h C(x)dx \quad (25)$$

$$C_t = \int_{h+1}^{P_t} C(x)dx \quad (26)$$

$C_h$ and $C_t$ determine the shape of C(x).

Using Eq. (16) in combination with the Eqs. (6), (8), (10), (11) and (14), we can write:

$$T = T(P_c, C_t, C_e, P_z) = \frac{P_c^2}{P} + \frac{C_t^2}{C} + \frac{C_e^2}{C} - \frac{P_z^2}{P} \qquad (27)$$

When $\frac{P_c^2}{P} + \frac{C_t^2}{C} + \frac{C_e^2}{C} > \frac{P_z^2}{P}$, T>0, which means a positive academic trace. When $\frac{P_c^2}{P} + \frac{C_t^2}{C} + \frac{C_e^2}{C} \le \frac{P_z^2}{P}$, T≤0, which means that the aggregated contributions do not add up sufficiently. In general, the sign of this T-value provides us with a first measure for the scientific quality of a document set under study.

Let us consider solving the first-order differentials of Eq. (27). When P and C are much larger than $P_c$, $C_e$ and $P_z$, P and C can be considered as constants, then we obtain:

$$\frac{\partial T}{\partial P_c} = \frac{2P_c}{P} > 0 \qquad (28)$$

$$\frac{\partial T}{\partial C_e} = \frac{2C_e}{P} > 0 \qquad (29)$$

$$\frac{\partial T}{\partial P_z} = -\frac{2P_z}{P} < 0 \qquad (30)$$

Since all second derivatives ($\frac{\partial^2 T}{\partial X^2} = const.\ [X = P_c, P_z, C_e]$) are constants, we are not able to decide for maximum or minimum values of $P_c$, $C_e$ and $P_z$. However, T tends to increase when $P_c$ or $C_e$ increases, and T decreases when $P_z$ increases. In other words, T does not only increase with $P_c$ = h, but the excess citations are also appreciated. A relatively large fraction of these can compensate for a large fraction of uncited publications in the aggregate.

## 6. Consistency

Rousseau and Ye (2012b) formulated the following independence axiom for any indicator f: If f(S) ≤ f(T) and the same type of basic steps are made to both sets S and T, then still f(S) ≤ f(T). Since P or C are independent indicators and X or Y are arithmetic combinations of P or C, respectively, X and Y also comply with the independence axiom.

Furthermore, since C-P is a simple consistent indicator (Rousseau & Ye, 2011), **Z=Y-X** is also a consistent indicator. Therefore, the performance matrix **V**=(X, Y, Z)$^T$ can be considered as a consistent two-dimensional measure. However, since T is a summation of the subsets $X_1$, $Y_2$ and $Z_3$, its consistency cannot be expected to hold under all conditions.

## 7. Discussion and conclusions

Using the h-value as a classifier, we introduce academic vectors: the row vector $\mathbf{X}=(X_1, X_2, X_3)$ indicates the relative publication distribution of the h-core, h-tail, and the uncited publications, while $\mathbf{Y}=(Y_1, Y_2, Y_3)$ denotes the relative citation distribution of the h-core, h-tail, and excess area, respectively. The column vector $(X_1, Y_1, Z_1)$ indicates the distribution of publications and citations in the h-core and the column vector $(X_2, Y_2, Z_2)$ does so in the h-tail. The column vector $(X_3, Y_3, Z_3)$ reflects the excess citations minus uncited publications, and row vector $\mathbf{Z}=(Z_1, Z_2, Z_3)$ represents citations minus publications in the three segments of h-core, h-tail, and uncited publications.

The performance matrix $\mathbf{V=(X,Y, Z)}^T$ constructs a unique two-dimensional measure for academic achievements and the academic trace of this matrix tr(**V**) provides a unique indicator for total academic achievements. As there may be a linear relation between **X** and **Y**, **V** can be at best two-dimensional; but never be more than two dimensions. Except perhaps for $X_3$ (uncited publications), larger values of $\{X_i, Y_i, Z_i\}$ indicate improvements in academic achievements.

The trace T compares citation with publication distributions like apples with oranges by providing both with a price in a single framework. Using this metaphor, one may consider the prices as fractions of the total number of publications and citations, respectively. Citations to publications in the h-tail thus have a different value from citations in the h-core or excess citations. The trace first aggregates the publications in the h-core with the citations in the h-tail, but then adds the excess citations as a fraction minus the fraction of uncited publications. The subtraction of the latter is perhaps the most debatable element of this indicator in evaluation research.

In terms of data collection, only five independent numbers $(P, P_z, h, C, C_h)$ are needed, although there are nine numbers that denote the multi-dimensional meanings in the matrix. All relevant data are available from WoS as summary statistics. The I3-like indicator in this core-tail framework of the h-index provides a conceptual link between these two indicators, and introduces new academic metrics, characterized by vectors **(X, Y, Z)**, the performance matrix $\mathbf{V=(X, Y, Z)}^T$ as well as the academic trace T=tr(**V**). The new indicator is well-grounded in mathematics, and contributes a useful, versatile, and easy-to-compute tool for the measurement and assessment of publication and citation profiles and can thus stimulate further studies.

**Acknowledgements**
Fred Ye acknowledges the National Natural Science Foundation of China Grant No 71173187 and National Social Science Foundation of China Major Key Project 12&ZD221 for financial supports. We thank Helen F. Xue for the data collection and Ronald Rousseau for pointing us to the difference between I3 versus I3X and I3Y, as well as anonymous reviewers for helpful comments and suggestions.

## Appendix: original data

**Table A1. Data for journals in two years 2009 and 2010.**

| Journal | P | $P_c$=h | $P_z$ | C | $C_h$ |
|---|---|---|---|---|---|
| *MIS Quart* | 88 | 20 | 6 | 1133 | 575 |
| *J Informetr* | 105 | 18 | 5 | 1132 | 574 |
| *J Am Med Inform Assn* | 247 | 24 | 26 | 2243 | 810 |
| *Annu Rev Inform Sci* | 24 | 7 | 5 | 159 | 129 |
| *J Inf Technol* | 69 | 8 | 18 | 227 | 107 |
| *Int J Comp-Supp Coll* | 45 | 10 | 9 | 276 | 170 |
| *Inform Manage-Amster* | 99 | 13 | 8 | 662 | 251 |
| *J Comput-Mediat Comm* | 93 | 12 | 10 | 583 | 284 |
| *Inform Syst Res* | 87 | 13 | 10 | 572 | 272 |
| *J Am Soc Inf Sci Tec* | 487 | 20 | 138 | 2404 | 712 |
| *Inform Syst J* | 56 | 9 | 11 | 274 | 140 |
| *Scientometrics* | 425 | 17 | 45 | 2247 | 435 |
| *MIS Q Exec* | 41 | 6 | 13 | 119 | 60 |
| *J Assoc Inf Syst* | 69 | 8 | 12 | 286 | 128 |
| *Libr Inform Sci Res* | 85 | 10 | 29 | 263 | 128 |
| *J Health Commun* | 193 | 12 | 38 | 776 | 219 |
| *Telecommun Policy* | 131 | 10 | 29 | 475 | 147 |
| *Int J Inform Manage* | 159 | 11 | 64 | 569 | 186 |
| *Eur J Inform Syst* | 99 | 9 | 10 | 365 | 126 |
| *Int J Geogr Inf Sci* | 175 | 13 | 32 | 795 | 241 |
| *J Strategic Inf Syst* | 43 | 7 | 10 | 175 | 97 |
| *Gov Inform Q* | 161 | 13 | 57 | 642 | 269 |
| *J Manage Inform Syst* | 89 | 10 | 21 | 368 | 168 |
| *J Inf Sci* | 97 | 9 | 16 | 391 | 124 |
| *J Knowl Manag* | 132 | 10 | 14 | 548 | 172 |
| *Inform Soc* | 80 | 6 | 33 | 155 | 53 |
| *Inform Process Manag* | 121 | 9 | 27 | 432 | 150 |
| *Soc Sci Comput Rev* | 72 | 8 | 23 | 224 | 101 |
| *J Doc* | 133 | 7 | 62 | 273 | 85 |
| *Serials Rev* | 93 | 4 | 63 | 64 | 18 |
| *J Med Libr Assoc* | 170 | 8 | 86 | 287 | 76 |
| *Online Inform Rev* | 193 | 10 | 105 | 360 | 128 |
| *Health Info Libr J* | 96 | 6 | 30 | 235 | 91 |
| *Learn Publ* | 132 | 6 | 80 | 178 | 72 |
| *Res Evaluat* | 77 | 6 | 25 | 186 | 65 |
| *Coll Res Libr* | 155 | 5 | 109 | 124 | 51 |
| *Libr Quart* | 74 | 4 | 49 | 71 | 29 |
| *Inform Res* | 169 | 3 | 153 | 23 | 10 |
| *Portal-Libr Acad* | 91 | 4 | 61 | 83 | 36 |
| *Inform Organ-Uk* | 27 | 5 | 4 | 68 | 32 |
| *Inform Technol Peopl* | 39 | 5 | 10 | 83 | 37 |
| *Data Base Adv Inf Sy* | 43 | 4 | 18 | 60 | 28 |
| *Libr Resour Tech Ser* | 68 | 5 | 35 | 77 | 31 |
| *Aslib Proc* | 79 | 6 | 36 | 146 | 65 |
| *J Scholarly Publ* | 67 | 3 | 45 | 50 | 20 |
| *Inform Technol Dev* | 49 | 5 | 20 | 86 | 40 |
| *Soc Sci Inform* | 56 | 4 | 22 | 80 | 24 |
| *J Acad Libr* | 237 | 6 | 155 | 206 | 46 |
| *J Libr Inf Sci* | 83 | 4 | 60 | 69 | 28 |
| *Rev Esp Doc Cient* | 61 | 3 | 42 | 40 | 14 |
| *Libr Cult Rec* | 95 | 2 | 79 | 21 | 4 |
| *Ethics Inf Technol* | 65 | 6 | 23 | 124 | 48 |
| *Libr Hi Tech* | 148 | 6 | 98 | 140 | 50 |
| *J Glob Inf Manag* | 31 | 5 | 8 | 74 | 33 |
| *Scientist* | 688 | 4 | 629 | 100 | 29 |
| *Electron Libr* | 214 | 7 | 128 | 224 | 71 |
| *Libr Collect Acquis* | 51 | 3 | 32 | 52 | 29 |
| *Online* | 215 | 3 | 196 | 30 | 10 |
| *Knowl Man Res Pract* | 73 | 5 | 25 | 123 | 40 |
| *Malays J Libr Inf Sc* | 42 | 4 | 22 | 46 | 19 |
| *Aust Acad Res Libr* | 103 | 4 | 84 | 46 | 19 |
| *Prof Inform* | 174 | 5 | 98 | 152 | 41 |
| *Libr Trends* | 87 | 3 | 44 | 74 | 14 |
| *Knowl Organ* | 65 | 3 | 45 | 33 | 9 |
| *Interlend Doc Supply* | 78 | 4 | 17 | 131 | 20 |
| *Program-Electron Lib* | 92 | 4 | 63 | 59 | 17 |
| *Aust Libr J* | 214 | 2 | 200 | 26 | 9 |
| *Libr J* | 8595 | 3 | 8561 | 47 | 15 |
| *Libri* | 55 | 3 | 29 | 45 | 14 |
| *Inform Technol Libr* | 64 | 3 | 57 | 47 | 12 |
| *Ref User Serv Q* | 310 | 3 | 280 | 48 | 13 |
| *Can J Inform Lib Sci* | 35 | 2 | 25 | 15 | 6 |
| *Restaurator* | 38 | 3 | 20 | 29 | 11 |
| *Inform Dev* | 65 | 3 | 45 | 38 | 13 |
| *Inform Technol Manag* | 34 | 3 | 16 | 31 | 9 |
| *Perspect Cienc Inf* | 134 | 2 | 119 | 18 | 4 |
| *Afr J Libr Arch Info* | 29 | 2 | 23 | 11 | 6 |
| *Investig Bibliotecol* | 65 | 1 | 61 | 4 | 1 |
| *Transinformacao* | 40 | 1 | 34 | 7 | 2 |
| *Z Bibl Bibl* | 120 | 2 | 114 | 8 | 4 |
| *Econtent* | 323 | 1 | 313 | 11 | 2 |
| *Libr Inform Sc* | 25 | 1 | 23 | 2 | 1 |
| *Inform Soc-Estud* | 79 | 1 | 73 | 7 | 2 |
| *Nature* | 5121 | 192 | 1834 | 182649 | 66109 |
| *Science* | 4955 | 171 | 1427 | 159648 | 52076 |
| *PNAS* | 8438 | 115 | 372 | 212651 | 18871 |

**Table A2. Data for two universities (2012) and two authors (2003-2012).**

| **Subject** | **P** | **$P_c$=h** | **$P_z$** | **C** | **$C_h$** |
|---|---|---|---|---|---|
| Univ Heidelberg | 4715 | 21 | 3149 | 5220 | 996 |
| Univ Hamburg | 1949 | 19 | 1257 | 3185 | 1243 |
| Leydesdorff L | 141 | 27 | 23 | 2183 | 1331 |
| Ye FY | 25 | 5 | 9 | 72 | 51 |